\documentclass[aps,twocolumn,amsmath,amssymb,showpacs,prb]{revtex4}
\usepackage{epsf}
\usepackage{graphicx}

\newcommand{\qt}{{\, \scriptstyle \otimes }\,}
\begin{document}
  \title{Superconducting Proximity Effect through a Magnetic Domain Wall}
  \author{Alexander Konstandin$^1$, 
  Juha Kopu$^{1,2}$, 
  and Matthias Eschrig$^1$, }
  \affiliation{$^1$Institut f\"ur Theoretische Festk\"orperphysik,
    Universit\"at Karlsruhe, 76128 Karlsruhe, Germany \\
    $^2$Low Temperature Laboratory,
    Helsinki University of Technology, FIN-02015 HUT, Finland }
  \date{July 20, 2005}   
\begin{abstract}
We study the superconducting proximity effect
in a superconductor-ferromagnet-superconductor (SFS) heterostructure,
containing a domain wall in the ferromagnetic region.
For the ferromagnet we assume an alloy with
an exchange splitting of the conduction bands
comparable to the superconducting gaps.
We calculate the modification of the density of states in 
the center of the domain wall
as a result of the proximity effect.
We show that the density of states is
sensitive to domain wall parameters due to
triplet-pairing correlations created in vicinity of the domain wall.
We present a theoretical tool which in a very effective way
enables retaining the {\it full } spatially dependent 
spin-space structure of the problem.
\end{abstract}

\maketitle

{\it Introduction:}
Most promising candidates for mesoscopic devices with novel
functionality are hybrid structures containing superconducting
elements. The key phenomenon that controls the behavior of such
systems is the proximity effect. When a superconducting
material is placed in contact with a normal metal (N), the superconducting
pair correlations leak over to the normal-metal side, changing its
conduction properties in the vicinity of the separating
interface. Quite similarly, the properties on the superconducting side
are also changed (the energy gap $\Delta_0$ is suppressed) due to the
contact to a normal metal. An alternative but equivalent way of
thinking about the proximity effect is through Andreev-reflection \cite{and}
processes: an incoming electron from the normal side is transmitted
together with another one as a Cooper pair into the superconducting
side. This phase-coherent electron-hole conversion results in a
nonzero pair amplitude in the normal metal. 

In the diffusive limit, the correlations relating
to an incident electron with an energy $E$ (the range of energies
being set by the temperature $T$) above the chemical potential extend
a characteristic distance of
$    \xi_N =\sqrt{D/E} $
into the normal metal;\cite{Kogan82}  here $D$ is the diffusion constant in N.
If the extent of the N region is finite, another energy scale,
$E_T \sim D/L^2$,
enters the problem; $L$ denotes the width of N. This
so-called Thouless energy has associated with it one of the generic
features of diffusive superconductor-normal metal 
heterostructures, the minigap: the density of states in the normal
metal develops a gap around the chemical potential in a manner similar
to a superconductor (S) but with a smaller magnitude. 

If the normal conductor is replaced by a ferromagnet (F), a multitude of
new effects arise due to the emergence of yet another energy scale,
that of the exchange splitting $J$ of the two spin bands. 
Both on the theoretical\cite{vol,kad,hal,rad,fom,bar,hue,morten,Buzdin05}
and on the experimental,\cite{kon,rya,gui,gir,gu,aum,gee,frolov,beckmann} side
interest has grown recently in the rich physics of such systems.
One source for new behavior is that, in the case with a singlet
superconductor, the induced pair amplitude in the ferromagnet is
oscillatory.\cite{buz} 
However, the exchange splitting
also gives rise to dephasing which, in turn, results in the decay
of induced correlations over a characteristic distance
$\xi_J=\sqrt{D/(E+J)}$.\cite{Bulaevskii85}
Unfortunately, since $J$ is of the order of the Fermi
energy $E_F$~in typical ferromagnetic metals, this distance is very
short.
Still, experimental
indication of the oscillatory behavior has been obtained in thin
ferromagnetic layers and, relevant to the present paper, in weakly
ferromagnetic alloys with $J \ll E_F$. \cite{kon,rya} 

Another question of current interest in SF proximity systems is the 
role of equal-spin triplet correlations.\cite{ber,mat} If created, 
e.g. near magnetic inhomogeneities, such
correlations would not be affected by the exchange splitting but could
penetrate considerably longer distances into F.\cite{ber}
Finally, the importance of domain walls 
has also been stressed for the Andreev conductance.\cite{cht,mel}.

\begin{figure}[b]
    \centerline{\epsfxsize=0.30\textwidth{\epsfbox{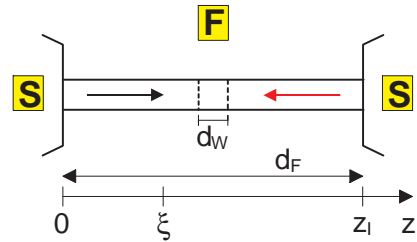}}}
    \caption{SFS structure with two magnetic domains oriented along the 
      $z$ axis and separated by a domain wall of width $d_W$; $d_F$ denotes
      the length of the F region and $\xi=\sqrt{D/\Delta_0}$ 
      is the superconducting coherence length.}
    \label{fig_SFS_ROT}
\end{figure}
In this paper we study an SFS structure, shown schematically in
Fig.~\ref{fig_SFS_ROT}, in equilibrium. 
The ferromagnetic region consists of two domains with magnetizations
oriented in opposite directions. The domains are separated by a domain
wall, where the magnetization rotates continuously between the
asymptotic values.  While varying in direction, the magnitude
$J$ is assumed constant throughout the F region. 
We show, that the local density of states (LDOS) in the F
region is strongly modified by the presence of the domain wall. 
In particular, we show that it can be 
very sensitive to the thickness of the domain wall in a certain parameter
region.

\newcommand{\kel}[1]{\check{\bf #1}}
{\it Basic equations:}
Proximity effect is a spatially inhomogeneous phenomenon.
An appropriate theoretical tool to treat such a problem
is the quasiclassical theory of superconductivity,
\cite{eil,lar} 
which in its diffusive version has been formulated by Usadel.\cite{usdl} 
In equilibrium, the physical information is
contained in the retarded Green functions $\hat{G}({
z},E)$.  Here, we assume spatial dependence in the coordinate $z$ only,
and $E$ denotes the energy as measured from the chemical potential.
The 4$\times$4 matrix structure, arising from 
particle-hole and spin degrees of freedom, is denoted by the
hat ( $\hat{}$ ) accent,
\begin{equation} 
\label{gl_green3}
    \hat{G}=\!
    \left( \begin{array}{cc} {\cal G} & {\cal F} \\ \tilde{\cal F} & \tilde {\cal G}
      \end{array} \right)\!.
\end{equation}
The off-diagonal elements determine the superconducting pair amplitude.
Quantities denoted
with the ``tilde'' are related to those without one through
$ \tilde{\cal A}(z,E)={\cal A}(z,-E^\ast)^\ast$.
All the elements in Eq.~(\ref{gl_green3}) are 2$\times$2 spin matrices:
e.g. ${\cal G}={\cal G}_{\alpha\beta}$ with $\alpha,\beta=\{\uparrow,
\downarrow\}$.
The Green functions satisfy the Usadel equation,\cite{usdl}
\begin{equation} 
\label{gl_usdl}
    \left[ E \hat{\tau}_3 -
    \hat{\Delta} - \boldsymbol{J} \cdot \hat{\boldsymbol{\sigma}},
    \hat{G} \right]_\qt +\frac{D}{\pi} \frac{{\rm d}}{{\rm d}z} \left( \hat{G} 
	\qt \frac{{\rm d } }{{\rm d}z} \hat{G}
    \right) = \hat{ 0}, 
\end{equation}
where the symbol $\qt $ denotes matrix multiplication, and $[\hat A,\hat B]_\qt
=\hat A \qt \hat B - \hat B \qt \hat A$.
In writing Eq.~(\ref{gl_usdl}) we have followed
the standard way to describe ferromagnetic materials 
through a spin-dependent energy shift,\cite{Bulaevskii85}
which 
has the form
$
E\hat\tau_3 \rightarrow E\hat\tau_3-\boldsymbol{J} \cdot \hat{\boldsymbol{\sigma}}.
$
Here, $\tau_3$ denotes the third Pauli-matrix in Nambu space, 
the vector $\boldsymbol{J}$ denotes the effective
exchange field of the ferromagnet, and $\hat{\Delta}$ is the
superconducting order parameter (appropriate for weak-coupling
spin-singlet pairing). 
The components of the vector 
$\hat{\boldsymbol{\sigma}}$ and the order parameter are given by
\begin{equation}
\hat{\sigma}_i=\left( \begin{array}{cc} \sigma_i & 0 \\ 0 & 
\sigma_i^\ast \end{array} \right),
\qquad
\hat{\Delta}=
\left( \begin{array}{cc} 0 & \Delta \\ \Delta^\ast & 0
\end{array} \right)\qquad
\end{equation}
where $\sigma_i$ are Pauli spin matrices, $i=x,y,z$,
and $\Delta=\Delta_0 i\sigma_y$. The above
procedure is appropriate for describing situations for which
$J\ll E_F$,
which holds e.g. for the 
ferromagnetic alloys used in Refs.~[\onlinecite{kon,rya}].
In writing Eq.~(\ref{gl_usdl}), we have chosen
the normalization according to
\begin{equation}
\label{norm}
\hat {G} \qt \hat {G}=-\pi^2 \hat {1}.
\end{equation}

{\it Riccati parameterization:}
The spin-dependent nature of SF proximity systems
calls for a formulation of the quasiclassical theory that retains the
full spin-space structure, especially in studying situations where the
exchange-field-orientation varies in space (such as in a domain
wall). Within the Eilenberger theory a very convenient
formulation already exists, \cite{esc} employing the so-called Riccati
parameterization. \cite{scho,nag} 
The extension of this method to the Usadel theory was achieved only recently,
\cite{Eschrig04} and has been applied to non-equilibrium situations,
\cite{Cuevas05}
and to FSF-systems with homogeneous magnetizations.\cite{Lofwander05}
Here we demonstrate its usefulness by applying it to a SFS system 
with a spatially changing magnetization in a domain wall, a case where the
conventional so-called $\theta $-parameterization\cite{belz} is not applicable.
The spin-dependent Riccati parameterization,\cite{esc}
\begin{eqnarray} 
\label{ricc1}
    \hat{G}\!&=& -i\pi \hat{N} 
	\qt \!\!\left(\!\begin{array}{cc}
     (1+\gamma \qt \tilde{\gamma}) & 2\gamma
    \\ -2\tilde{\gamma} &
    -(1+\tilde{\gamma} \qt \gamma)
	\end{array}\!\right),
\end{eqnarray}
with 
\begin{equation}
\label{ricc3}
     \hat{N}=\!\!
	\left(\!\!\begin{array}{cc}
     (1-\gamma\qt\tilde{\gamma})^{-1} & 0 
    \\ 0 &
      (1-\tilde{\gamma}\qt\gamma)^{-1}
	\end{array}\!\!\right)
\end{equation}
automatically accounts for the
normalization (\ref{norm}), which is essential for practical
numerical calculations.
It is enough to determine one 2$\times$2 matrix in spin space, $\gamma$.
The other, $\tilde \gamma $, follows from the above-mentioned (fundamental)
symmetry.
The transport equation for $\gamma $ follows from Eq.~(\ref{gl_usdl}), and reads\cite{Eschrig04}
\begin{eqnarray}
\label{ussolve}
\frac{{\rm d}^2\gamma }{{\rm d}z^2}
&+&\left(\frac{{\rm d} \gamma }{{\rm d}z}\right)
\qt \frac{\tilde{\cal F}}{i\pi}\qt
\left(\frac{{\rm d} \gamma }{{\rm d}z}\right) 
=\frac{i}{D}\bigg[\gamma\qt{\Delta}^\ast\qt\gamma
\bigg. \nonumber \\
&&\bigg.-(E-\boldsymbol{J} \cdot \boldsymbol{\sigma})\qt
\gamma-\gamma\qt (E+ \boldsymbol{J} \cdot \boldsymbol{\sigma}^\ast)
-\Delta 
\bigg], \qquad
\end{eqnarray}
Here, the expression for $\tilde{\cal F}$ is obtained by comparing
Eq.~(\ref{gl_green3}) with Eqs.~(\ref{ricc1})--(\ref{ricc3}).

{\it Boundary conditions:}
Additionally, boundary conditions
are required for the different interfaces of the system. 
Such conditions have been formulated by Nazarov.\cite{naz} 
The outer surfaces ($z=z_{o}$) of the superconductors 
are assumed to border to an insulating
region, and the appropriate condition is 
$\partial_z \hat G(z_o, E)=0$, {\it i.e.}
\begin{equation} 
\label{bc1}
\frac{d\gamma}{dz}(z_{o},E)=0.  
\end{equation} 
On the other hand, the two inner SF interfaces ($z=z_i^S$ for the
S side, $z=z_i^F$ for the F side) are assumed
in the following transparent. 
The corresponding boundary conditions are in this case
$\hat{G}(z_i^S,E)=\hat{G}(z_i^F,E)$, 
$\sigma_S \partial_z \hat {G} (z_i^S,E)=
    \sigma_F \partial_z \hat {G} (z_i^F,E)$, leading to 
\begin{eqnarray} 
\label{bc2}
\nonumber
\gamma(z_{i}^S,E)&=& \gamma(z_{i}^F,E), \\
\sigma_S\frac{d\gamma}{dz}(z_{i}^S,E) &=&
\sigma_F\frac{d\gamma}{dz}(z_{i}^F,E),  
\end{eqnarray} 
where $\sigma_{S}$ and $\sigma_F$ refer to the conductivities
of S and F, respectively.
For simplicity, we have assumed $\sigma_S=\sigma_F$,
implying the continuity of the derivative at the
interface. With the boundary conditions (\ref{bc1}) and (\ref{bc2}),
we have solved Eq.~(\ref{ussolve}) numerically by an iterative
procedure (relaxation method) in the entire SFS system.  

\begin{figure}[t]
    \includegraphics[width=0.45\textwidth]{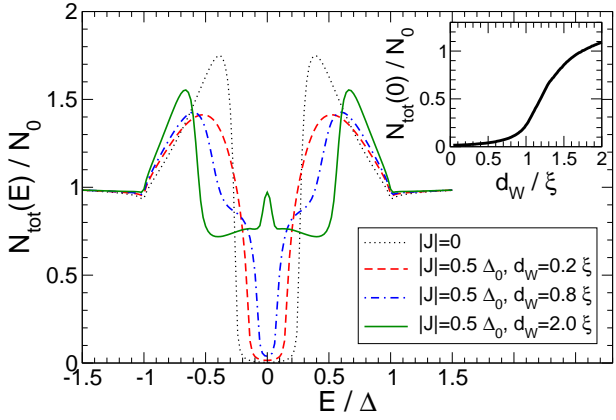}\vspace{0.1cm}
    \caption{LDOS as a function of energy for the system of
    Fig.~\ref{fig_SFS_ROT}, calculated in the middle of the F region
    for different widths $d_W$ of the domain wall. Here,
    $J=0.5~\Delta_0$, $d_F=2.0~\xi$. For comparison, the dotted
    line shows the corresponding dependence in the normal-metal case
    ($J=0$).
    The inset shows for $J=0.5~\Delta_0$, 
    the value of the LDOS at the chemical potential
    ($E=0$) as a function of the domain-wall width $d_W$.
    }  \label{Fig2}
\end{figure}

{\it SFS system with domain wall:}
We apply the outlined theory to study the SFS structure of
Fig.~\ref{fig_SFS_ROT} in equilibrium. Lengths are given in units of
the superconducting coherence length, $\xi=\sqrt{D/\Delta_0}$.
The spin-singlet superconductors are chosen to have the
same gap magnitude. The contact areas at the SF interfaces
are assumed to be small enough, so that any gap suppression can be neglected.
The two
superconducting regions and the intermediate ferromagnet are taken to
have fixed lengths of $d_S=5\xi$ and $d_F=2\xi$.
We model the domain wall by a varying 
direction of ${\bf J}=(J_x,J_y,J_z)$ (keeping
the magnitude $J=|{\bf J}|$ constant), with $J_y=0$ and
\begin{equation}
 \left( \begin{array}{c} 
J_x\\ J_z \end{array} \right)=
J \left( \begin{array}{c} 
\cos \theta(z) \\ \sin \theta (z)
\end{array}
\right), 
\; \theta (z)= -\arctan \frac{z-z_0}{d_W}.
\end{equation}
Here, $d_W$ is an effective domain wall width parameter.
In the following we study the influence of the width $d_W$ of a domain wall
centered in F ($z_0=d_F/2$) on the density of states in the center
of the domain wall ($z=z_0$).

Knowing $\gamma(z,E)$ from the solution of Eq.~(\ref{ussolve})
with boundary conditions (\ref{bc1}) and (\ref{bc2}),
the quasiclassical Green function and the (total) LDOS 
\begin{equation}
\label{ldos}
N_{tot}(z,E)=-\frac{N_0}{2\pi}~{\rm Im~Tr}~{\cal G}(z,E),
\end{equation}
is determined via Eqs.~(\ref{ricc1})-(\ref{ricc3}); 
$N_0$ is the normal-state
density of states, and Tr denotes the spin trace. 

An important characteristic of the value of the LDOS in superconductor-normal metal
proximity systems is the minigap: the density of states in the
normal-metal region shows a gap of width $E_g < \Delta_0$ induced by
proximity to a superconductor. 
The energy $E_g$ can be thought
of as that of the lowest-energy Andreev bound state in a finite
normal-metal layer.  This convenient physical picture can easily be
extended to single-domain ferromagnets. 
The corresponding spin-dependent energy shift of the quasiparticle and the
Andreev-reflected quasihole by $\pm J$ leads to a reduction of
the energy of the lowest-lying
bound state, and correspondingly of the minigap, 
from the expression for a normal metal by $J$, 
vanishing altogether when $J\ge E_{g,J=0}$.
This picture is confirmed by our numerical calculations.

In the inhomogeneously magnetized case of Fig.~\ref{fig_SFS_ROT}, the
above picture is modified. The effect of the domain wall on the 
LDOS is summarized in Fig.~\ref{Fig2},
which shows $N_{tot}$ as a function of energy for different domain-wall
widths $d_W$. Although the value of $J=0.5\Delta_0$ is 
here larger than the value
of the normal state minigap $E_g\approx 0.25 \Delta_0$ (as seen from the
dotted curve in Fig.~\ref{Fig2} for $J=0$), the minigap is reduced to zero
only for larger domain wall widths $d_W\approx 2\xi$.
For the smallest width $d_W=0.2\xi$ the minigap is only reduced by about
40\%. The additional states which fill the minigap with increasing
domain wall width are due to spin triplet correlations, which are
sensitive to the {\it direction} of ${\bf J}$.
Our calculations show that the influence of equal-spin
pairing components created by the domain wall increases.
This is reflected by the appearance of
additional Andreev bound states inside the gap, modifying the LDOS.  
The relative importance of the triplet correlations depends on $J$
and $d_W$: as clearly manifested by Fig.~\ref{Fig2},
the efficiency of the triplet-inducing
mechanism grows with increasing $d_W$. 
The inset of Fig.~\ref{Fig2} shows the value of the LDOS at $E=0$
as a function of $d_W$. The interesting observation here is that the 
LDOS at the chemical potential is very sensitive to the domain wall
width when the latter one is comparable to $\xi$. 

  \begin{figure}[t]
  \centerline{\includegraphics[width=0.45\textwidth]{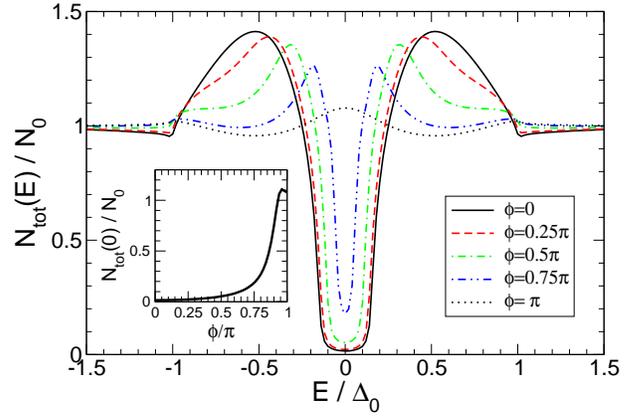}}
  \vspace{0.1cm}
  \caption{LDOS as a function of energy for several phase
  differences $\phi$ between the superconductors, calculated in the middle 
  of F. Here, $J=0.5~\Delta_0$, $d_F=2.0~\xi$.
  The width of the domain wall is $d_W=0.2~\xi$.
  The inset shows the corresponding LDOS at the chemical potential ($E=0$) 
  as a function of $\phi$.
  }
  \label{Fig3} \end{figure}

\begin{figure}
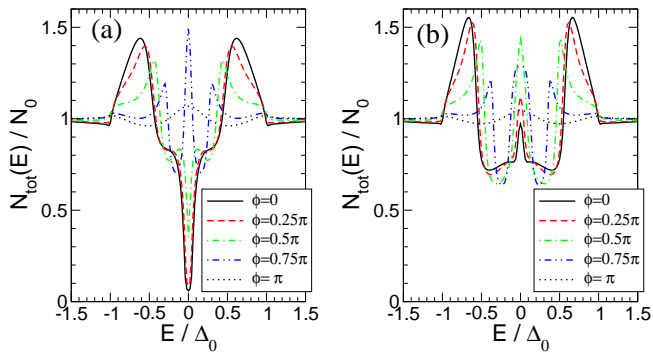
 \vspace{0.3cm}		
    \centerline{\includegraphics[width=0.23\textwidth]{Fig4a.eps}
    \hfill
    \includegraphics[width=0.23\textwidth]{Fig4b.eps} }
    \caption{LDOS as a function of energy for several phase
  differences $\phi$ between the superconductors, calculated in the middle
  of F, and for $J=0.5~\Delta_0$, $d_F=2.0~\xi$. 
  The width of the domain wall is in (a) $d_W=1.0~\xi$, 
  and in (b) $d_W=2.0~\xi$.}
  \label{Fig4} 
  \end{figure}

Finally, with a view towards studying the possible effects of
the domain walls on the supercurrent flowing in an SFS structure,
we have studied the LDOS in the case where there is a 
phase difference $\phi$ between the two superconductors. 
This phase difference adds to the one accumulated by the quasiparticles
and quasiholes in the ferromagnetic region and, thus, modifies the
spectrum of Andreev bound states.
Figures \ref{Fig3} and \ref{Fig4} present
the LDOS in the middle of the F region for three domain walls
with different widths. 
In the inset of Fig. \ref{Fig3} we also show
the zero-energy LDOS for a domain wall of width $d_W=0.2\xi$ 
as a function of the phase difference. 
As can be seen in Fig.~\ref{Fig4}, by a possible tuning of the
domain wall width $d_W$ one can always find a region of strongest
sensitivity for a given phase difference $\phi$ and vice versa.
This increases the possibilities of controlling the zero energy
density of states in the domain wall.
The rich structure exhibited by these results could
easily result in highly nontrivial behavior of the transport
current both in equilibrium and nonequilibrium situations. 

{\it Conclusions:}
We have studied numerically the LDOS in a heterostructure
consisting of a ferromagnetic alloy sandwiched
between two singlet superconductors. 
We find strong modifications of the LDOS
caused by the presence of a domain wall. 
As only triplet superconducting correlations are sensitive to the
direction of the exchange field, the strong variations in the 
LDOS result from the presence of triplet correlations
induced by the spatially varying magnetization.
We also find a strong dependence of the density of states in the
domain wall on a possible phase difference 
between the superconducting order parameters, giving an additional tool
to control its value. This motivates future studies of the interplay of a
supercurrent and the domain wall (Josephson effect).
We hope that the variety of features observed in
our calculations motivates further experimental research on
proximity systems involving weak ferromagnets. 

This work was supported by Deutsche Forschungsgemeinschaft within the
Center for Functional Nanostructures.

\end{document}